\documentclass[preprint2]{aastex}

\slugcomment{submitted to ApJ}

\shorttitle{X-ray sources in M31}
\shortauthors{Kaaret}

\begin{document}

\title{A Chandra High-Resolution Camera Observation of X-Ray Point
Sources in M31}

\author{Philip Kaaret}

\affil{Harvard-Smithsonian Center for Astrophysics,
Cambridge, MA 02138}

\begin{abstract}

We present results from a 47~ks observation of the Andromeda galaxy,
M31, using the High-Resolution Camera of the Chandra X-Ray
Observatory.  We detect 142 point sources spanning three orders of
magnitude in luminosity, from $L_X = 2\times 10^{35} \rm \, erg \,
s^{-1}$ to $L_X = 2\times 10^{38} \rm \, erg \, s^{-1}$ in the
0.1-10~keV band.  The X-ray source location accuracy is better than
$1\arcsec$ in the central regions of the galaxy.  One source lies
within $1.3\arcsec$ of SN 1885 but does not coincide with the UV
absorption feature identified as the supernova remnant.  However, there
is an optical transient, which is likely an optical nova, at the
location of the X-ray source.   There is a weak source, $L_X \sim 4
\times 10^{36} \rm \, erg \, cm^{2} \, s^{-1}$, coincident with the
nucleus of M31, and 14 sources coincident with globular clusters.

Our observation has very high efficiency down to luminosities of
$1.5\times 10^{36} \rm \, erg \, s^{-1}$ for sources within $5\arcmin$
of the nucleus.   Comparing with a ROSAT observation made 11 years
earlier, we find that $0.46 \pm 0.26$ of the sources with $L_X > 5
\times 10^{36} \rm \, erg \, s^{-1}$ are variable.  We find no evidence
for X-ray pulsars in this region, indicating that the population is
likely dominated by low-mass X-ray binaries.  The source density radial
profile follows a powerlaw distribution with an exponent of $1.25 \pm
0.10$ and is inconsistent with the optical surface brightness profile. 
The x-ray point source luminosity function is well fitted by a
differential broken powerlaw with a break at a luminosity of
$(4.5^{+1.1}_{-2.2}) \times 10^{37} \rm \, erg \, s^{-1}$.  The
luminosity function is consistent with a model of an aging population
of X-ray binaries.

\end{abstract}


\section{Introduction}

Study of X-ray sources in external galaxies is important for
understanding the formation history and population statistics of X-ray
binaries and other X-ray sources both in external galaxies and in our
own.  Such studies will help us understand the evolutionary history of
X-ray binaries, should provide information on the star formation
history (e.g.\ White \& Ghosh 1998), and may be important in estimating
the rate of merging objects (e.g.\ Bethe \& Brown 1999) critical for
determining the rate of gravitational wave events.

The nearby, bright spiral galaxy M31 offers an excellent site for such
studies.  The distance of the galaxy is well known, making luminosity
estimation from flux measurement straight forward, the galaxy is
sufficiently inclined so that sources can be reliably located within
its morphology, and it is relatively nearby (780~kpc) so that many
sources can be detected.  The sub-arcsecond resolution of the Chandra
X-ray Observatory (CXO; Weisskopf 1988) permits individual X-ray
sources to be discerned even in crowded regions of the galaxy and
enables measurement of highly accurate positions.  Such position
information is critical in finding unique optical and radio
counterparts to the sources.

Here, we report on a deep observation of the core of M31 made with the
Chandra High Resolution Camera (HRC; Murray et al.\ 1997).  The HRC
offers the best spatial resolution available for the X-ray study of
M31.  We describe the observation and our analysis in \S 2.  We discuss
the variability in \S 3 and  source identifications including a source
very near, but not coincident with SN 1885, in \S 4.  We consider the
spatial and luminosity distributions of the population of X-ray point
sources in \S 5 and describe a model which relates the luminosity
distribution to the age of the X-ray binary population.  Finally, we
conclude in \S 6 with comments on the implications of our results for
understanding the formation history of X-ray binaries.

\begin{figure*}[t] \centerline{\epsscale{1.6} \plotone{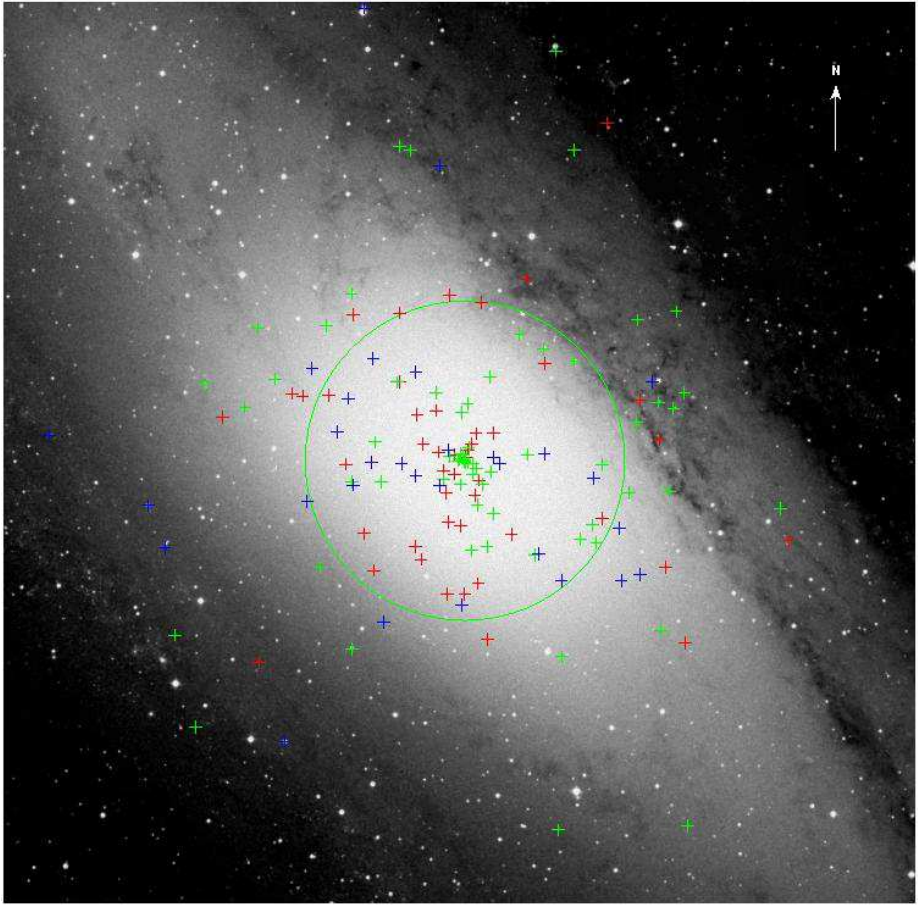}}
\caption{Optical image of M31 from the Digital Sky Survey with
positions of Chandra sources superimposed.  The sources are color-coded
according to their luminosity: red for $L_X \le 2 \times 10^{36} \rm \,
erg \, s^{-1}$, magenta for $2 \times 10^{36} \rm \, erg \, s^{-1} <
L_X \le 2 \times 10^{37} \rm \, erg \, s^{-1}$, and blue for $L_X > 2
\times 10^{36} \rm \, erg \, s^{-1}$.  The green circle indicates the
sources used in the analysis of group properties.  It is centered on
the nucleus and and has a radius of $5\arcmin$.  The arrow is
$2\arcmin$ long and points north. \label{bigpic}} \end{figure*}

\begin{figure}[t] \centerline{\epsscale{1.0} \plotone{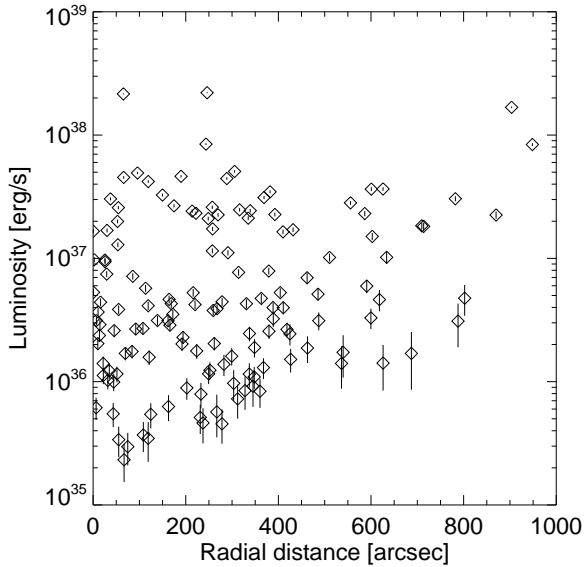}}
\caption{Luminosity versus radial distance from the nucleus for X-ray
point sources in M31. \label{lum_r}}   \end{figure}

\section{Observation and Source List}

M31 was observed with the Chandra X-Ray Observatory (CXO; Weisskopf
1988) using the High Resolution Camera (HRC; Murray et al.\ 1997) in
imaging mode and the High-Resolution Mirror Assembly (HRMA; van
Speybroeck et al.\ 1997) for 46777~s of useful exposure time beginning
at  23:49:20 UTC on 2001-Oct-31 (ObsID 1912).  The data were
reprocessed using the Chandra geometry file released on 2001-Oct-22
(telD1999-07-23geomN0004.fits) and then standard event processing and
filtering procedures, from the Chandra Interactive Analysis of
Observations software package ({\it CIAO}) v2.2.1, were applied to
produce a level-2 event list.  A light curve from the full detector was
constructed to search for times of high background.  The count rate
appears uniform through out the observation.

Sources were extracted from three images which were produced from the
level-2 event list. The first covered the central $400\arcsec \times
400\arcsec$ region with a resolution equal to the HRC electronic pixels
($0.13\arcsec$).   The second covered the central $800\arcsec \times
800\arcsec$ region with a resolution of 2 HRC pixels.  The third image
covered the full HRC field of view with a resolution of 4 HRC pixels.
The Chandra point spread function (PSF) degrades off axis.  Use of
three nested images allowed us to adequately sample the Chandra PSF at
each off-axis angle without requiring processing of excessively large
images.

We constructed a source list for each image using {\it wavdetect}
\citep{freeman02}, the wavelet-based source detection routine in {\it
CIAO}.  We merged the three source lists. We removed duplicates and
extracted the position and region parameters for each source  with a
detection significance above $4\sigma$ from the highest resolution
image in which it appeared.  The source regions are ellipses with radii
equal to 3 times the 50\% encircled energy radii calculated by {\it
wavdetect} from the events detected for each source.  The merged source
regions were visually inspected.  We manually reduced the sizes of
overlapping source regions to reduce the overlap. This was important
for the three sources near the nucleus of M31 and two pairs of
sources:  CXOM31 J004302.9+411522 with CXOM31 J004303.2+411528, and
CXOM31 J004255.3+412558 with CXOM31 J004253.4+412549.  The fluxes
quoted for these sources may be slightly underestimated.  

We extracted counts for each source region.  We estimated the
background for each source using a circular region with a radius 3
times the major axis of the source ellipse and excluding regions twice
the size (both major and minor axes doubled) of the source regions. 
The exposure was calculated for an average photon energy of 1~keV. 
Relative to the value on-axis, the exposure decreases by 10\% at
$6\arcmin$ off-axis, by 30\%  at $12\arcmin$ off-axis, and by 42\% at
the source furthest off-axis.  We translated the count rates to photon
fluxes using the exposure, then to energy fluxes in the 0.1-10~keV band
assuming an absorbed thermal bremsstrahlung spectrum with a temperature
of 2 keV and a column density equal to the Galactic column towards M31
of $6.66 \times 10^{20} \rm \, cm^{-2}$, and finally to equivalent
luminosities at a distance of 780~kpc \citep{stanek98}.  The conversion
to luminosity is uncertain due to the lack of spectral information,
e.g.\ use of a temperature of 5~keV would increase the luminosities by
a factor of 1.5, while use of 1~keV would decrease them by 0.7, and 
use of a powerlaw spectrum with a photon index of 2 would increase them
by 1.2.  For comparison with \citet{tf91} and \citet{primini93}, who
quoted luminosities in the 0.2-4~keV band for a thermal bremsstrahlung
spectrum with a temperature of 5 keV and a distance of 690~kpc, our
values should be multiplied by 0.76.

Table~\ref{table:src} includes sources with estimated luminosities as
low as $2.3\times 10^{35} \rm \, erg \, s^{-1}$.  However, the
detection threshold is not uniform across the HRC field of view.  The
degradation of the HRMA point spread function far off-axis causes the
photons from sources far off-axis to be spread over a large physical
area on the detector.  This increases the background within the source
region and worsens the detection sensitivity.  Fig.~\ref{lum_r}
shows the source luminosity as a function of angular distance from the
Chandra aimpoint. We note that the spatially varying diffuse emission
near the nucleus could also be of concern for the detection threshold. 
However, the diffuse emission is strong only in the central parts of
the galaxy where the point spread function is narrow and, hence, adds
relatively few photons within the source regions.  Sources with
luminosities above $7 \times 10^{35} \rm \, erg \, s^{-1}$ are
efficiently detected out to a distance of $300\arcsec$ from the
aimpoint. The detection threshold rises to slightly below  $2 \times
10^{37} \rm \, erg \, s^{-1}$ at $700\arcsec$ from the aimpoint, and
rapidly becomes worse at larger radii.

To check the astrometry of the Chandra observation, we compared the
Chandra source positions to the 2mass infrared point source catalog
\citep{2mass_ref}.  We selected the 2mass catalog because it contains
more point sources near the central regions of M31 than do optical
astrometry catalogs, such as USNO or Tycho, and because it has good 
absolute astrometry with an accuracy of $\lesssim 0.2\arcsec$
\citep{2mass_ref}.  Within $6\arcmin$ of the Chandra aimpoint, where
the point spread function is narrow and the source positions are well
defined, we find 6 coincidences within $0.5\arcsec$ excluding sources
within $5\arcsec$ of the nucleus.  There are 107 Chandra sources and
110 2mass sources in this same region.  The expected number of chance
coincidences within $0.5\arcsec$ is 0.02.  Therefore it is unlikely
that more than one of the 6 coincidences is due to chance.  We
calculated the average coordinate shift for the 6 sources and used this
to correct the Chandra position.  The magnitude of the shift was
$0.2\arcsec$.  The positions in Table~\ref{table:src} include this
correction.  After the shift, all 6 sources are coincident within
$0.4\arcsec$ with an average magnitude of displacement of
$0.15\arcsec$.

\section{Variability}

We searched for variability on long time scales ($\sim 10$ years) by
comparing the Chandra fluxes with fluxes measured in a ROSAT
observation made during 1990 \citep{primini93} and on short times
scales ($\sim 10^4 \rm s$) by looking at light curves extracted from
our Chandra data.  We also searched for periodic signals.

\begin{figure}[t] \centerline{\epsscale{1.0} \plotone{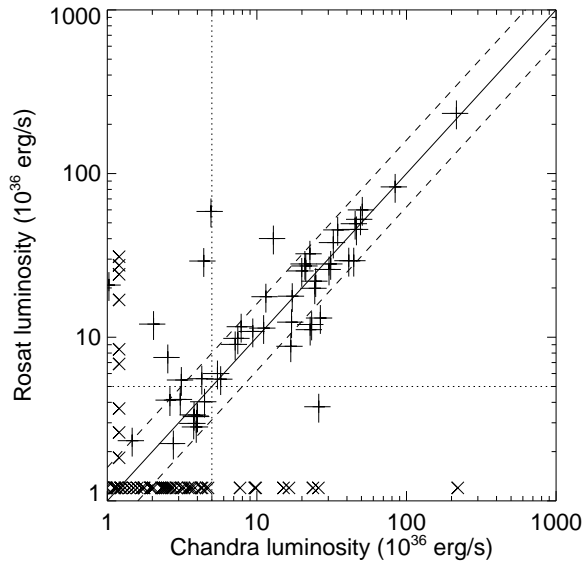}}
\caption{Comparison of source luminosities measured by Rosat in 1990
and by Chandra in 2001. The crosses represent sources with position
matches within $3\arcsec$.  The X's are sources without matches.  The
solid line indicates equal luminosities. The dashed lines indicate a
factor of 1.6 change in luminosity.  The dotted lines indicate the
threshold luminosity of $5\times 10^{36} \rm \, erg \, s^{-1}$. 
\label{vrosat}}   \end{figure}

\subsection{Long Time Scales}

\citet{primini93} derived a source list from a ROSAT High-Resolution
Imager (HRI) observation of the central region of M31 made on 25-28
July 1990.  Because the sensitivity and point spread function of both
ROSAT and Chandra decrease as one goes off-axis, we restricted our
comparison to sources with $7.5 \arcmin$ of the nucleus of M31 (which
was close to the aimpoint for both observations).  Comparing the source
lists, we searched for matches and found an average offset of
$-0.37\arcsec$ in RA and $+0.89\arcsec$ in DEC. After adding these
offsets to the ROSAT positions, we found 53 coincidences within
$3\arcsec$.  To compare with our Chandra results, we multiplied the
ROSAT luminosities luminosities by a factor of 1.31 to correct for the
different distance, spectral model, and spectral range assumed. To
allow for calibration errors and uncertainties induced by the lack of
spectral information in either observation, we assign a systematic
error of 20\% on each luminosity.

Fig.~\ref{vrosat} shows a comparison of the ROSAT and Chandra source
luminosities.  There are many more Chandra sources below a luminosity
of $5\times 10^{36} \rm \, erg \, s^{-1}$ which lack a counterpart than
there are ROSAT sources.  This likely indicates incompleteness in the
ROSAT sample at low luminosities, and we restrict our comparison to
sources which appeared above this luminosity in at least one of the
observations.  We consider a source to be ``constant'' if the
luminosity varies by less than a factor of 1.6 between the two
observations.  This is a reasonably conservative criterion.

We find 55 distinct sources in the two observations.  Of these, 29 have
constant luminosity (within a factor of 1.6) between the two
observations.  Hence, the fraction of sources which are variable is
$0.47 \pm 0.22$.  This number is relatively insensitive to changes in
the criteria, e.g.\ changing the threshold for luminosity variations
from 1.6 to 2.0 changes the variable fraction to $0.42 \pm 0.22$ which
is consistent within the errors.  Considering only sources with
$5\arcmin$ of the nucleus and taking a luminosity variation threshold
of 1.6 gives a variable fraction of $0.46 \pm 0.26$.  The fraction of
sources found here to be variable on time scales of $\sim 10$ years is
similar to the number found to be variable on time scales less than 2
years by \citet{kong02}.

\begin{figure}[t] \centerline{\epsscale{0.8} \plotone{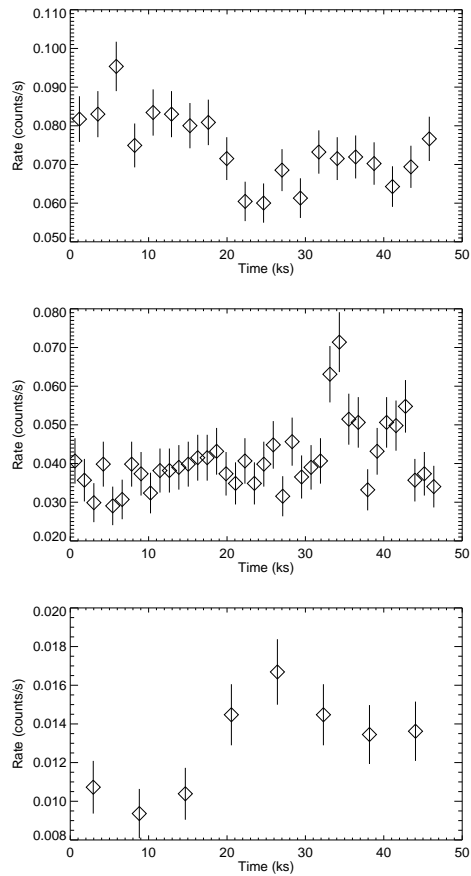}}
\caption{Light curves for three sources showing variability.  The top
panel is for CXOM31 J004222.9+411535, the center for CXOM31
J004252.5+411854, and the bottom for CXOM31 J004318.9+412016.
\label{lc}}   \end{figure}

\subsection{Short Time Scales}

We searched for variability within the Chandra observation by
extracting the photon arrival times for each source and then comparing
them to the distribution expected for a constant source with the same
average flux using a KS test.  No background subtraction was performed.
Three sources were found to be variable at a confidence level above
99.9\%.  Their light curves are shown in Fig.~\ref{lc}.  These three
sources were also found to be variable on time scales of months to
years by \citet{kong02}.  CXOM31 J004222.9+411535 shows an irregular
decline over the observation.  CXOM31 J004318.9+412016 shows an
increase in flux by 50\% over a time scale of 10~ks.  Such behaviors
are common to X-ray binaries. 

CXOM31 J004252.5+411854 shows a flare with a duration of $\gtrsim 2000
\rm \, s$, a peak flux of $8 \times 10^{37} \rm \, erg \, s^{-1}$, and
a fluence (total energy including persistent emission) of $1.8 \times 
10^{41} \rm \, erg $ within the first 2000~s.  The duration of the
flare and the fluence may be substantially larger if the high flux
points following the flare are interpreted as part of the flare, i.e.\
the duration may be as long as 10~ks and the fluence as large as $5
\times  10^{41} \rm \, erg $.  The flare properties are similar to
those of the recently discovered ``superbursts'' \citep{cornelisse00}
which may be due to unstable burning of $^{12}$C near the neutron star
surface \citep{cumming01}.  Also, the average luminosity of the source
is within the luminosity range, 0.1--0.3 of the Eddington luminosity
for a $1.4 M_{\odot}$ neutron star, of the known superburst sources
\citep{wijnands01sb}.  If identification of the observed flare as a
superburst is correct, then CXOM31 J004252.5+411854 would be identified
as a neutron-star low-mass X-ray binary.

\subsection{Pulsations}

We conducted an accelerated search for periodic signals with periods in
the range 0.25--5000~s and allowing acceleration sufficient to cover
orbital periods down to 6~days for companion star masses up to $40
M_{\odot}$ \citep{wood91}.  The search range would permit detection of
pulsations from any of the known Be/X-ray binaries, except for
A0538-66.  Taking into account the large number of trials, we found no
signal significant at above the 99\% confidence level.  We note that
XMMU J004319.4+411759 found by \citet{osborne01} to exhibit 865~s
pulsations was not detected (even as unpulsed source) in our
observation.

Given the number of trials over accelerations, 143, and over
frequencies, $1.9 \times 10^{5}$, we can place a 99\% confidence upper
bound on the pulsed fraction for sinusoidal pulsations from each source
of $\eta = \sqrt(69/N)$, where $N$ is the number of photon detected
from the source \citep{buccheri87}. Concentrating on the sources within
$5\arcmin$ of the nucleus, we find 13 sources with more than 1000
counts for which the pulsed fraction upper bound is $\eta < 0.26$ and
22 sources with more than 500 counts for which the upper bound is $\eta
< 0.37$.

These upper bounds are below the pulsed fractions of many known
Be/X-ray binaries suggesting that the core of M31 is home to few, if
any, high-mass X-ray binaries.  Combined with the high fraction of
sources which are variable, this leads to the conclusion that the X-ray
point source population of the core of M31 is dominated by  low-mass
X-ray binaries (LMXBs) as previously suggested by other authors (e.g.
van Speybroeck et al.\ 1979).

\section{Source Identifications}

We searched for optical and infrared counterparts to the Chandra
sources.  In this section, we discuss some of the counterparts found.

\subsection{Globular clusters and infrared sources}

Globular clusters are well known to contain X-ray sources, in both our
galaxy \citep{clark75} and M31 (e.g.\ di Stefano et al.\ 2001), so we
searched for coincidences between M31 globular clusters
\citep{battistini80,magnier93,barmby01} and our Chandra sources. 
Infrared emission is common from globular clusters, so we also included
sources from the 2mass catalog \citep{2mass_ref} as a check on the
astrometry of the globular cluster catalogs.  We list coincidences
within $2.5\arcsec$ of Chandra HRC sources with 2mass sources or
globular clusters in Table~\ref{glob_ir}. In the cases where a
particular globular cluster appeared in more than one catalog, we list
the entry from the catalog of \citet{barmby01} where available and
\citet{magnier93} otherwise.  The cluster position is taken from
catalog containing the closest coincidence.  We exclude the nucleus
from the list of 2mass sources as it is discussed in a later section.

We consider there to be strong evidence for an association of an X-ray
source with a globular cluster when the positions coincide within
$1\arcsec$.  There are 14 such coincidences.  We expect only 0.4 chance
coincidences within $1\arcsec$.  Of these 14 sources, 8 have 2mass
counterparts within $2.5\arcsec$ and the average radial offset between
the 2mass and Chandra sources is $0.4\arcsec$.

Several of the remaining coincidences may be due to chance.  We expect
3.3 chance coincidences with separations between $1\arcsec$ and
$3\arcsec$ compared with 6 coincidences found.  However, some of the
coincidences may represent true associations with the displacement
overestimated due to errors in the cluster or X-ray position.  The
latter is particularly relevant for  CXOM31 J004215.9+410115, CXOM31
J004301.2+413017, and CXOM31 J004337.1+411443 which lie more than
$10\arcmin$ from the Chandra aimpoint.  Follow-up observations centered
on their locations would be useful to confirm or deny the associations.

Only two of the X-ray sources with 2mass counterparts lack globular
cluster counterparts.  These two infrared sources are slightly dimmer,
but have colors consistent with those of the 2mass sources with
globular cluster and X-ray counterparts.  CXOM31 J004221.5+411419 lies
within $0.3\arcsec$ of a globular cluster candidate (S5 15) identified
by \citet{wirth85}.  It is likely the same object as the X-ray source
2E 0039.6+4057 seen with Einstein. CXOM31 J004210.2+411510 is likely
the same source identified as \#6 by \citet{primini93} and as CXOM31
J004210.1+411509 by \citet{kong02}.  We are aware of no globular
cluster, or other, counterpart to the source.

\begin{figure}[t] \centerline{\epsscale{1.0} \plotone{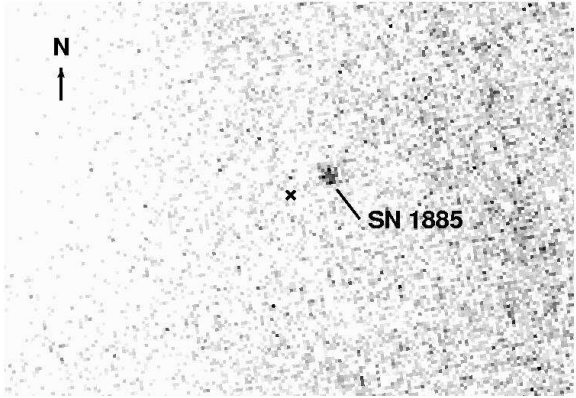}}
\caption{Near UV image of SN 1885 following \citet{fesen99}. The X
marks the position of CXOM31 J004243.1+411604 when the UV and X-ray
images are aligned as described in the text.  The arrow at the upper
left points North and has a length of $1\arcsec$. \label{sn1885zoom}}  
\end{figure}

\begin{figure}[t] \centerline{\epsscale{1.0} \plotone{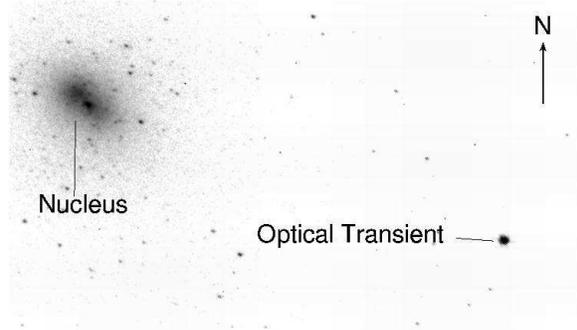}}
\caption{HST image of a bright optical transient coincident with CXOM31
J004243.1+411604. The transient is located in the lower, right part of
the image and is coincident with CXOM31 J004243.1+411604.  The nucleus
of M31 is at the left.  The arrow at the upper right points North and
has a length of $2\arcsec$. \label{opttrans}}   \end{figure}

\subsection{Source near SN 1885}

The X-ray source CXOM31 J004243.1+411604 lies close to the position
of the historical supernova SN 1885 \citep{barnard98,deV85}. No X-ray
source has been previously reported at this position, even in
observations made with Chandra as late as 10 June 2001 \citep{kong02},
only 143 days before our observation.  \citet{fesen99} found a near-UV
absorption disk in a Hubble Space Telescope (HST) image of the central
region of M31.  To determine whether the Chandra source is the
counterpart of SN 1885, we extracted the WFPC2 data of \citet{fesen99}
from the HST archive and repeated the processing described in their
paper to obtain a median filtered image at $3915\AA$, similar to that
shown in Fig.~1 of their paper.  The absolute astrometry of HST is
limited by the accuracy of the guide star positions which have a
typical uncertainty of $0.5\arcsec$ rms ($0.8\arcsec$ at 90\%
confidence).  The accuracy of relative astrometry is much better.  

A bright UV source is present about $1\arcmin$ SW of the nucleus and
lies near the Chandra source CXOM31 J004241.3+411524; both sources are
coincident with the globular cluster identified as MIT 213
\citep{magnier93}.  Assuming that the UV source is, indeed, the
counterpart of the Chandra source allows us to align the UV and X-ray
images.  Fig.~\ref{sn1885zoom} shows the UV image with the position of
CXOM31 J004243.1+411604 superimposed.  The Chandra source lies
$1.3\arcsec$ from the center of the UV absorption disk, indicating that
CXOM31 J004243.1+411604 is not the counterpart of SN 1885.  The
angular displacement at the distance of M31 corresponds to a
(projected) physical displacement of 4.9~pc.  A velocity close to
40,000~km/s would be required for an object to move such a distance
between 1885 and 2001.  This is much higher than the velocity of any
known pulsar.  Hence, interpretation of the Chandra source as a neutron
star ejected from the SN 1885 is excluded.

We examined other HST images of M31 and found an optical transient
source at the position of CXOM31 J004243.1+411604 in a WFPC2 image (HST
data set U2LG020, originally taken by \citet{lauer93} to study the
nucleus of M31).  Fig.~\ref{opttrans} shows an image obtained with the
F300W filter on 19 June 1995 including the region around SN 1885.  We
aligned this image with the X-ray image by registering it with other
HST images \citep{garcia01} including the bright UV/optical source
identified with CXOM31 J004241.3+411524 above.  A bright point source
is apparent which is coincident with CXOM31 J004243.1+411604 within
$0.15\arcsec$.  This is compatible with the accuracy expected from our
alignment procedure. Discovery of this optical counterpart strengths
our belief that identification of the bright UV source and CXOM31
J004241.3+411524 is correct (even though the X-ray position is
$1.5\arcsec$ from the cataloged globular cluster position) and that the
Chandra source is not the counterpart of SN 1885.  The HST source is
the brightest object other than the nucleus in the HST image.  Using
the standard HST calibrations, we estimate the flux is $3.5 \times
10^{-16} \rm \, erg \, cm^{-2} \, s^{-1} \AA^{-1}$, equivalent to an ST
magnitude of 17.5.  The brightness is comparable to that of optical
novae detected in M31 \citep{arp56,shafter01} and we suggest that the
transient is most likely an optical nova.

Some novae have been detected as supersoft X-ray sources which turn on
within a few years after the nova outburst  \cite{kahabka97}.  Such
sources have very soft spectra, typically blackbody spectra with
temperatures of 20--80~eV.  A blackbody spectrum with a temperature of
80~eV would imply an unabsorbed luminosity of $9 \times 10^{36} \rm \,
erg \, s^{-1}$ in the 0.1-10~keV band for HRC count rate of CXOM31
J004243.1+411604.  Lower temperatures would require higher
unabsorbed luminosities to produce the same observed count rate.  This
luminosity would be consistent with the properties of the known
supersoft sources associated with novae outbursts \citep{orio01}.
Spectral observations of CXOM31 J004243.1+411604 would be of great
interest to determine if it is, indeed, a super soft source.

\subsection{Sources near the nucleus}

Three Chandra/HRC sources lie within $5\arcsec$ of the nucleus of M31. 
Of these, only CXOM31 J004244.3+411608 overlaps the bright optical
nucleus apparent in HST images of the core of M31 \citep{garcia01}. 
This source is well detected with 152 net counts, but is relatively dim
with a flux of $(1.44 \pm 0.12) \times 10^{-5} \rm \, photons \,
cm^{-2} \, s^{-1}$.  The corresponding luminosity is $3.6 \times
10^{36} \rm \, erg \, cm^{2} \, s^{-1}$ in the 0.1--10~keV band
assuming a thermal bremsstrahlung spectrum with a temperature $kT = \rm
2 \, keV$; the true luminosity may be somewhat different if the true
spectrum has a different form (e.g.\ a powerlaw spectrum with photon
index of 2 would give a luminosity of $4.5 \times 10^{36} \rm \, erg \,
cm^{2} \, s^{-1}$).  This luminosity is a factor of $\sim 1000$ higher
than that measured for Sgr A* \citep{baganoff01}.  Because of the high
density of Chandra sources near the nucleus, we cannot exclude the
possibility that the X-ray emission does not arise from the nucleus of
M31.  Hence, the X-ray emission from CXOM31 J004244.3+411608 should be
taken as an upper bound on the emission from the nucleus of M31.

\section{Group properties}

Here, we consider the group properties of the X-ray point sources in
the core of M31. We concentrate our analysis on sources within a
$5\arcmin$ radius of the nucleus and with luminosities above 
$1.5\times 10^{36} \rm \, erg \, s^{-1}$, corresponding to 63 net
counts near the aimpoint.  Our detection efficiency is very high given
these limits.  Hence, our sample should be essentially complete. The
sample also corresponds to a well-defined physical area covering the
bulge of M31 within 230~pc of the nucleus, assuming a distance of
780~kpc.

\begin{figure}[t] \centerline{\epsscale{1.0} \plotone{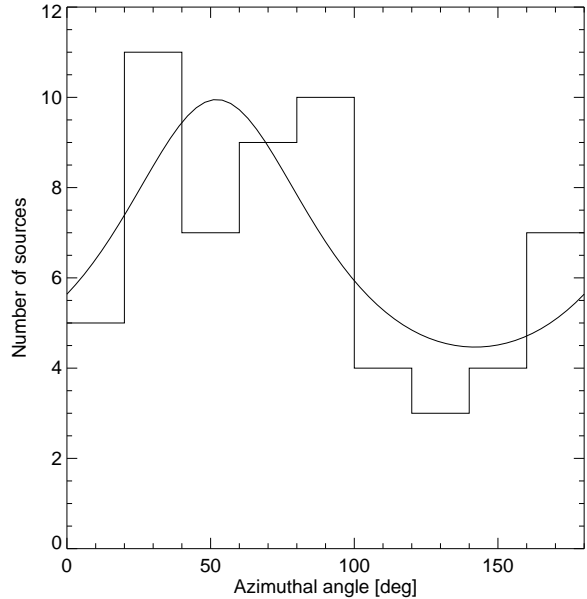}}
\caption{Azimuthal distribution of X-ray point sources in M31.  The
curve is the best fit ellipsoidal profile. \label{azdist}}  
\end{figure}

\begin{figure}[t] \centerline{\epsscale{1.0} \plotone{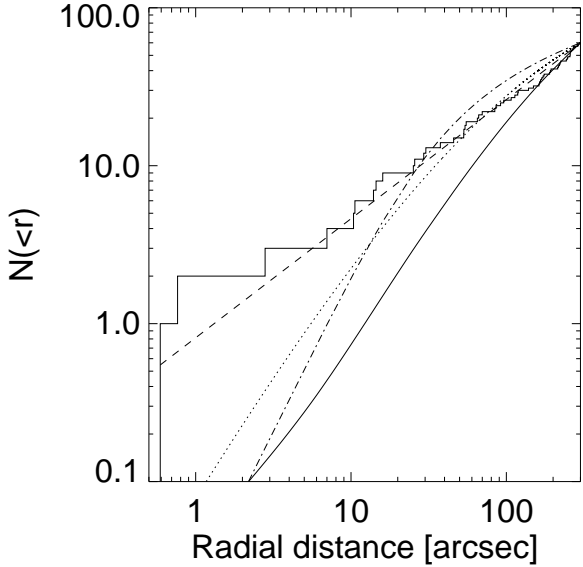}}
\caption{Cumulative number of X-ray point sources in M31  versus radial
distance from the nucleus. The stepped solid line is the X-ray data. 
The smooth solid line is the optical profile of \citet{kent87} scaled
to match the X-ray data at large radii.  The dashed line is a powerlaw
fit, the dash-dotted line a King model, and the dotted line is a de
Vaucoulers profile. \label{dens_cum}}   \end{figure}

\subsection{Spatial Distribution of Sources} \label{sec:spat}

The azimuthal distribution of the X-ray point sources within a
$5\arcmin$ radius of the nucleus and with luminosities above 
$1.5\times 10^{36} \rm \, erg \, s^{-1}$ is shown in Fig.~\ref{azdist}.
A fit of an ellipsoidal distribution gives a position angle $\theta =
52\arcdeg \pm 13\arcdeg$ measured from north to east and an ellipticity
$\varepsilon = 0.33 \pm 0.11$.  \citet{kent83,kent87} measured the
distribution of optical surface brightness for M31.  He found that the
position angle is near $52\arcdeg$ within $5\arcsec$ of the nucleus,
shifts to a value as low as $42\arcdeg$ at a radius of of $10\arcsec$,
and then returns to $\sim 50\arcdeg$ at $5\arcmin$. The ellipticity is
also non-monotonic being relatively high, $\varepsilon = 0.3$ at small
radii, decreasing to $\varepsilon = 0.1$ at intermediate radii near
$30\arcsec$, and then increasing to $\varepsilon \sim 0.3$ at
$5\arcmin$.  The limited number of X-ray point source do not permit
study of the azimuthal distribution as a function of radius, but the
general properties are similar to the optical distribution.  The X-ray
distribution is aligned, within errors and the variations in the
optical distribution, with the optical distribution and has a similar
ellipticity.  This suggests that the X-ray and optical distributions
are physically related.

Fig.~\ref{dens_cum} shows the cumulative radial distribution of the
same set of X-ray point sources.  The X-ray point source distribution
differs significantly from the optical surface brightness profile;
there are many more X-ray sources near to the nucleus than would be
expected from scaling to the optical profile matched to the X-ray
profile at large radii.  To constrain the shape of the distribution, we
fit the data to a powerlaw form, a King model, and a de Vaucoulers
profile. The fitting was done using an unbinned maximum entropy method
\citep{cash79}.  As is apparent in the figure, the powerlaw form
provides a fully adequate fit, while the King model and the de
Vaucoulers profile significantly underestimate the number of sources at
small radii.  The best fit powerlaw exponent is $1.25 \pm 0.10$.

\begin{figure}[t] \centerline{\epsscale{1.0} \plotone{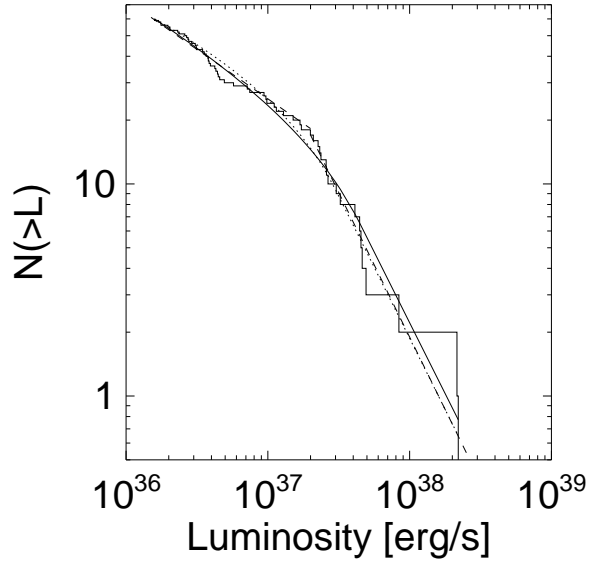}}
\caption{Cumulative luminosity functions of X-ray point sources in
M31.  The solid curve is for sources within $5\arcmin$ of the nucleus
and the short-dashed curve is for all sources. The long-dashed curve is
a broken powerlaw fit to the solid curve. \label{xlffit}} \end{figure}

\subsection{X-ray luminosity distribution} \label{sec:xlf}

The cumulative X-ray point source luminosity function (XLF) for sources
with luminosities above $1.5\times 10^{36} \rm \, erg \, s^{-1}$ and
within a $5\arcmin$ radius of the nucleus are shown in
Figure~\ref{xlffit}.  The XLF shows a distinct break near $2-3\times
10^{37} \rm \, erg \, s^{-1}$ as has been reported in previous studies
of the X-ray population of the core of M31 \citep{primini93,shirey01}. 
The sample used to construct this XLF has high and uniform sensitivity
which minimizes problems arising from incompleteness or non-uniformity,
while still extending to luminosities an order of magnitude below the
break.

To compare with previous results, we fit the XLF with two powerlaw
functions: one for the range luminosity range $0.15-2 \times 10^{37}
\rm \, erg \, s^{-1}$ and one for $L_X >  2 \times 10^{37} \rm \, erg
\, s^{-1}$.  Using a minimum entropy fitting procedure appropriate for
our situation of high detection efficiency \citep{crawford70} and
fitting the unbinned luminosity data, we find slopes for the cumulative
distributions of $1.36 \pm 0.33$ for the higher luminosity range and
$0.47 \pm 0.09$ for the lower luminosity range.  The fit is shown as
the long-dashed curve in Fig.~\ref{xlffit}.  The slopes we obtain are
consistent with the results of \citet{primini93}, \citet{shirey01}, and
\citet{kong02}.  

We next fit the XLF using a broken powerlaw for the differential
luminosity function.  We note that the cumulative distribution
corresponding to the differential broken powerlaw has curvature below
its break luminosity and is not the same form as a cumulative broken
powerlaw.  The break luminosity for the differential broken powerlaw
lies above the luminosity where the eye would locate the break in the
cumulative XLF.  This distinction must be noted when comparing break
luminosities derived from cumulative, e.g.\ \citet{shirey01}, versus
differential broken powerlaws, e.g.\ \citet{sarazin00}.   We used an
unbinned maximum entropy technique \citep{cash79} and found that the
best fit break luminosity is $(4.5^{+1.1}_{-2.2}) \times 10^{37} \rm \,
erg \, s^{-1}$.  Roughly a factor of 2 higher than the break luminosity
quoted for the cumulative distribution above.  For comparison, we fit
the luminosity data from \citet{shirey01} with a differential broken
powerlaw (the data were kindly provided by R.\ Shirey and R.\ Soria)
and find a break luminosity of $(4.5 \pm 2.1) \times 10^{37} \rm \, erg
\, s^{-1}$ in good agreement with the break luminosity derived from our
Chandra data and roughly a factor of two higher than the break
luminosity of $2.5 \times 10^{37} \rm \, erg \, s^{-1}$ derived from
their visual inspection of their cumulative luminosity distribution.

The best fit index for the differential powerlaw below the break
luminosity is $1.30^{+0.13}_{-0.18}$ while that above is
$2.5^{+0.6}_{-0.5}$.  The corresponding cumulative powerlaw slopes
would be one less.  These values are in reasonable agreement with the
fits to the cumulative luminosity distribution.  The slope of the
differential low luminosity powerlaw is somewhat flatter than that of
the cumulative low luminosity powerlaw due to the curvature noted
above.

Finally, we fit the XLF with a simple model in which all of the x-ray
sources are members of a single population with uniform properties
except for luminosity and lifetime \citep{kilgard02}.  Our sample is
confined to the bulge of M31.  It is, most likely, predominately a
uniform population of low-mass X-ray binaries from the bulge of M31,
with minor contamination from other classes of sources.  In the
modeling, we assume that X-ray binaries are born with a birth
luminosity distribution given as a powerlaw, $b(L) \propto L^{-\alpha}$
at a rate which has been constant in time over an interval $t_S$.  For
simplicity, we assume that each source has a constant luminosity
through its lifetime which is $\tau = \eta \bar{M}_2 c^2 /\epsilon L$,
where $\eta$ is the efficiency for conversion of accreted matter to
radiation, $\bar{M}_2$ is the average mass of the companion stars, and
$\epsilon$ is the duty cycle of emission \citep{king01}.  

If we assume that $\eta$, $\epsilon$, and $\bar{M}_2$ are independent
of luminosity, then the differential luminosity distribution is
described by a broken powerlaw with the index for the high luminosity
powerlaw being steeper than that of the low luminosity powerlaw by one
\citep{kilgard02}.  The break luminosity is related to the model
parameters as $L_S = \eta \bar{M}_2 c^2 /\epsilon t_S$. The break
luminosity is then an indicator of the age of the X-ray binary
population.  We note that the difference between the upper and lower
powerlaw indexes for both the differential broken powerlaw fit are
consistent with 1.0 within errors.  Fitting with a differential broken
powerlaw in which the difference between the two indexes is fixed to
1.0, shown as the solid line in  Fig.~\ref{xlffit}, we find a fit which
is as good as  the general broken powerlaw fit given above.  The best
fit break luminosity is $(4.4^{+1.2}_{-2.2}) \times 10^{37} \rm \, erg
\, s^{-1}$ and the best fit slope for the differential low luminosity
powerlaw is $1.33^{+0.12}_{-0.20}$.

The difference between the upper and lower powerlaw indexes is one only
when $\eta$, $\epsilon$, and $\bar{M}_2$ are independent of the source
luminosity and there is no evolution of luminosity over the life of an
individual source.  More generally, we can relate the break luminosity
to the source age as $\tau \propto L^{-\beta}$.  In this case the
difference between the powerlaw indexes will be $\beta$ and the  break
luminosity will vary with the population age as $L_S \propto
t_S^{-\beta}$.  From our fitting results using the differential
powerlaw in which both the upper and lower indexes are free, we can
constrain $\beta = 1.2 \pm 0.6$.

\section{Conclusions}

Previous X-ray studies of M31 with pre-Chandra instruments indicated a
good correspondence between the X-ray and optical profiles
\citep{tf91}.  However, the good correspondence does not continue to
the small radii accessible with Chandra.  Instead, we find a powerlaw
profile with an exponent of $1.25 \pm 0.10$.  Powerlaw profiles with
exponents near 1 are well known indicators of core collapse in globular
clusters \citep{djorgovski86}.  The same should be true for other
stellar clusters such as the bulge of M31 \citep{quinlan96}.  If the
X-ray sources are predominately LMXBs which were formed in globular
clusters \citep{white02}, then the distribution may be a remnant of the
inward migration and disruption of globular clusters
\citep{tremaine75}.

The X-ray point source luminosity function (XLF) of the core of M31
shows a distinct break near $4.4 \times 10^{37} \rm \, erg \, s^{-1}$.
A break near this luminosity has been reported previously for M31
\citep{primini93,shirey01}, although our analysis shows that the break
luminosity found from fitting a broken powerlaw to the differential
distribution is roughly a factor of two higher than that found from
fitting to the cumulative distribution.  

In the context of the model presented above \citep{kilgard02}, the
break luminosity is related to the age of the X-ray point source
population.  Specifically, the X-ray population age $t_B = 1.3 \eta
\bar{M}_2 /\epsilon M_{\odot} \rm \, Gyr$.  For accreting neutron stars
in low-mass X-ray binaries, the efficiency $\eta \sim 0.1$ and the
average companion mass $\bar{M_2} \sim M_{\odot}$.  Good observational
constraints on the duty cycle of emission are not available.  The duty
cycles inferred for LMXBs in our Galaxy range from 1.0 to less than
0.01 \citep{wijnands01}.  It has also been suggested, based on binary
evolution models, that a population of LXMBs with low duty cycles is
present in elliptical galaxies \citep{piro02}.  The stellar population
near the nucleus of M31 is one of the reddest known, leading to age
estimates of 10~Gyr or older \citep{rich95}.  If the X-ray population
is of the same age, then $t_B \gtrsim 10 \rm \, Gyr$ and we would
estimate $\epsilon \sim 0.01$.  If the duty cycle were larger then the
compact objects would accrete a total mass exceeding the companion mass
in less than 10~Gyr.

\section{Acknowledgments}

I thank Michael Garcia for providing the HST mosaic image, Pauline
Barmby for processing the HST optical transient image, Frank Primini
for providing the ROSAT source list, all three of them and Dong-Woo
Kim, Albert Kong, Andrea Prestwich, Andreas Zezas, and Rick Harnden for
valuable discussions, Robert Shirey and Roberto Soria for providing
their XMM luminosity data, and the CXC team for their superb operation
of Chandra.  PK acknowledges partial support from NASA grant NAG5-7405
and Chandra grant number G01-2034X.  This publication makes use of data
products from the Two Micron All Sky Survey, which is a joint project
of the University of Massachusetts and the Infrared Processing and
Analysis Center/California Institute of Technology, funded by the
National Aeronautics and Space Administration and the National Science
Foundation.

\begin{deluxetable}{llrrcccl}
\tablecaption{Chandra HRC sources in M31 \label{table:src}}
\tablewidth{0pt}
\tablehead{
  \colhead{RA} & \colhead{DEC} &
    \colhead{S/N} & \colhead{Counts} & 
    \colhead{Flux} & 
    \colhead{Luminosity} \\
  \colhead{(J2000)} & \colhead{(J2000)} & 
  \colhead{} & \colhead{} & 
  \colhead{$\rm (10^{-6} \, photons \, cm^{-2} \, s^{-1})$} & 
  \colhead{$(10^{36} \rm \, erg \, s^{-1})$} }
\startdata
 00 41 50.46 & +41 13 37.8 &    6.3 &    49.8 &     5.7 $\pm$  2.3 &    1.4 \\
 00 41 51.78 & +41 14 37.7 &    6.5 &   118.6 &    13.3 $\pm$  2.5 &    3.3 \\
 00 42 07.34 & +41 04 43.0 &    4.6 &   144.7 &    19.2 $\pm$  5.4 &    4.8 \\
 00 42 07.67 & +41 10 26.6 &    5.9 &    49.8 &     5.7 $\pm$  2.1 &    1.4 \\
 00 42 07.77 & +41 18 15.1 &   55.8 &   665.2 &    69.0 $\pm$  2.9 &   17.0 \\
 00 42 09.15 & +41 20 47.8 &   16.4 &   196.9 &    20.9 $\pm$  2.2 &    5.2 \\
 00 42 09.61 & +41 17 45.3 &   20.9 &   214.4 &    22.1 $\pm$  1.9 &    5.5 \\
 00 42 10.28 & +41 15 10.4 &   18.7 &   156.4 &    16.1 $\pm$  1.5 &    4.0 \\
 00 42 10.91 & +41 12 48.6 &    8.6 &    57.2 &     6.1 $\pm$  1.3 &    1.5 \\
 00 42 11.74 & +41 10 49.0 &    9.3 &   111.2 &    12.4 $\pm$  2.0 &    3.1 \\
 00 42 11.96 & +41 16 49.4 &    9.6 &    52.0 &     5.3 $\pm$  1.0 &    1.3 \\
 00 42 12.14 & +41 17 58.4 &   13.9 &   100.5 &    10.3 $\pm$  1.4 &    2.5 \\
 00 42 13.15 & +41 18 36.7 &  122.4 &  1369.3 &   140.4 $\pm$  3.9 &   34.7 \\
 00 42 15.16 & +41 12 34.6 &   76.9 &   874.7 &    91.9 $\pm$  3.2 &   22.7 \\
 00 42 15.21 & +41 18 01.8 &    6.6 &    44.8 &     4.5 $\pm$  1.0 &    1.1 \\
 00 42 15.53 & +41 20 31.9 &   13.6 &   102.6 &    10.6 $\pm$  1.4 &    2.6 \\
 00 42 15.73 & +41 17 20.7 &   21.9 &   172.4 &    17.3 $\pm$  1.5 &    4.3 \\
 00 42 15.93 & +41 01 15.5 &   26.5 &  2067.6 &   339.2 $\pm$  9.0 &   83.8 \\
 00 42 17.01 & +41 15 08.5 &   36.9 &   310.7 &    31.2 $\pm$  1.9 &    7.7 \\
 00 42 18.36 & +41 12 24.1 &  114.9 &  1202.8 &   126.0 $\pm$  3.7 &   31.1 \\
 00 42 18.65 & +41 14 02.1 &  110.0 &   988.1 &   100.1 $\pm$  3.2 &   24.7 \\
 00 42 20.57 & +41 26 40.7 &    4.1 &    58.0 &     6.9 $\pm$  3.4 &    1.7 \\
 00 42 21.49 & +41 16 01.4 &  116.6 &   711.8 &    69.8 $\pm$  2.6 &   17.2 \\
 00 42 21.56 & +41 14 19.6 &    5.6 &    17.9 &     1.8 $\pm$  0.6 &    0.4 \\
 00 42 22.45 & +41 13 34.2 &   69.0 &   444.8 &    44.9 $\pm$  2.2 &   11.1 \\
 00 42 22.95 & +41 15 35.5 &  492.5 &  3469.5 &   340.1 $\pm$  5.8 &   84.0 \\
 00 42 23.17 & +41 14 07.7 &   27.9 &   159.2 &    15.9 $\pm$  1.3 &    3.9 \\
 00 42 25.15 & +41 13 40.9 &   15.4 &    82.5 &     8.2 $\pm$  1.0 &    2.0 \\
 00 42 26.06 & +41 19 15.0 &   32.6 &   183.9 &    18.1 $\pm$  1.4 &    4.5 \\
 00 42 26.11 & +41 25 51.0 &    9.4 &   165.2 &    18.7 $\pm$  3.7 &    4.6 \\
 00 42 28.21 & +41 10 00.8 &   55.7 &   610.4 &    66.4 $\pm$  2.9 &   16.4 \\
 00 42 28.30 & +41 12 23.3 &  227.8 &  1774.8 &   179.7 $\pm$  4.3 &   44.4 \\
 00 42 28.88 & +41 04 36.1 &   17.8 &   572.3 &    73.6 $\pm$  5.1 &   18.2 \\
 00 42 29.11 & +41 28 57.0 &    5.4 &   101.9 &    12.5 $\pm$  4.8 &    3.1 \\
 00 42 30.99 & +41 19 10.9 &    5.4 &    19.8 &     1.9 $\pm$  0.6 &    0.5 \\
 00 42 31.14 & +41 16 21.8 &  331.9 &  1369.8 &   131.3 $\pm$  3.6 &   32.4 \\
 00 42 31.25 & +41 19 38.9 &   78.5 &   476.7 &    46.5 $\pm$  2.2 &   11.5 \\
 00 42 32.08 & +41 13 14.6 &  147.9 &   945.0 &    94.0 $\pm$  3.1 &   23.2 \\
 00 42 32.75 & +41 13 11.2 &   34.3 &   171.8 &    17.1 $\pm$  1.3 &    4.2 \\
 00 42 33.89 & +41 16 20.0 &   57.1 &   161.4 &    15.4 $\pm$  1.2 &    3.8 \\
 00 42 34.16 & +41 21 49.5 &    5.5 &    31.8 &     3.2 $\pm$  0.9 &    0.8 \\
 00 42 35.21 & +41 20 06.0 &   30.3 &   157.0 &    15.3 $\pm$  1.3 &    3.8 \\
 00 42 36.62 & +41 13 50.4 &    7.7 &    22.9 &     2.2 $\pm$  0.5 &    0.6 \\
 00 42 38.58 & +41 16 03.9 & 1329.0 &  9219.4 &   872.8 $\pm$  9.1 &  215.6 \\
 00 42 39.54 & +41 14 28.7 &   78.3 &   239.5 &    23.2 $\pm$  1.5 &    5.7 \\
 00 42 39.57 & +41 16 14.5 &  260.5 &  1104.0 &   104.4 $\pm$  3.1 &   25.8 \\
 00 42 39.65 & +41 17 00.7 &    5.5 &    12.7 &     1.2 $\pm$  0.4 &    0.3 \\
 00 42 39.98 & +41 15 47.7 &  144.2 &   550.0 &    52.2 $\pm$  2.2 &   12.9 \\
 00 42 40.21 & +41 18 45.3 &   55.2 &   193.1 &    18.5 $\pm$  1.3 &    4.6 \\
 00 42 40.56 & +41 10 33.5 &    6.2 &    42.8 &     4.5 $\pm$  1.0 &    1.1 \\
 00 42 40.68 & +41 13 27.6 &   32.3 &   113.5 &    11.2 $\pm$  1.1 &    2.8 \\
 00 42 41.43 & +41 15 24.0 &   53.8 &   163.2 &    15.6 $\pm$  1.2 &    3.9 \\
 00 42 41.59 & +41 21 05.8 &   10.4 &    64.6 &     6.4 $\pm$  1.1 &    1.6 \\
 00 42 42.06 & +41 15 32.1 &   14.6 &    42.2 &     4.0 $\pm$  0.6 &    1.0 \\
 00 42 42.14 & +41 12 18.2 &    4.6 &    16.2 &     1.6 $\pm$  0.5 &    0.4 \\
 00 42 42.33 & +41 14 45.6 &   96.1 &   300.2 &    29.0 $\pm$  1.7 &    7.2 \\
 00 42 42.46 & +41 15 53.9 &  103.7 &   399.9 &    38.0 $\pm$  1.9 &    9.4 \\
 00 42 42.52 & +41 16 59.4 &    5.0 &    14.4 &     1.4 $\pm$  0.4 &    0.3 \\
 00 42 42.74 & +41 15 03.8 &    4.2 &     9.8 &     0.9 $\pm$  0.3 &    0.2 \\
 00 42 42.97 & +41 15 43.3 &   83.1 &   316.4 &    30.2 $\pm$  1.7 &    7.5 \\
 00 42 43.11 & +41 16 04.2 &   33.8 &   123.6 &    11.7 $\pm$  1.1 &    2.9 \\
 00 42 43.18 & +41 16 40.4 &   17.4 &    52.3 &     5.0 $\pm$  0.7 &    1.2 \\
 00 42 43.31 & +41 13 19.8 &   49.2 &   175.5 &    17.4 $\pm$  1.3 &    4.3 \\
 00 42 43.74 & +41 16 32.6 &  104.7 &   413.3 &    39.2 $\pm$  1.9 &    9.7 \\
 00 42 43.80 & +41 16 29.2 &   11.8 &    47.9 &     4.5 $\pm$  0.7 &    1.1 \\
 00 42 43.85 & +41 16 04.1 &   35.2 &   133.3 &    12.7 $\pm$  1.1 &    3.1 \\
 00 42 43.95 & +41 17 55.7 &   37.3 &   116.2 &    11.1 $\pm$  1.0 &    2.7 \\
 00 42 44.26 & +41 16 14.4 &    7.7 &    26.2 &     2.5 $\pm$  0.5 &    0.6 \\
 00 42 44.35 & +41 16 08.9 &   47.7 &   151.6 &    14.4 $\pm$  1.2 &    3.6 \\
 00 42 44.35 & +41 16 05.5 &  101.3 &   418.2 &    39.8 $\pm$  2.0 &    9.8 \\
 00 42 44.37 & +41 16 07.6 &  159.2 &   641.1 &    61.0 $\pm$  2.4 &   15.1 \\
 00 42 44.38 & +41 11 58.7 &   10.6 &    46.1 &     4.7 $\pm$  0.8 &    1.2 \\
 00 42 44.66 & +41 16 18.3 &   42.0 &   156.3 &    14.9 $\pm$  1.2 &    3.7 \\
 00 42 44.84 & +41 11 38.4 &  116.2 &   888.4 &    91.5 $\pm$  3.1 &   22.6 \\
 00 42 44.88 & +41 17 40.0 &   33.6 &   113.5 &    10.8 $\pm$  1.0 &    2.7 \\
 00 42 45.08 & +41 15 23.3 &   33.8 &   109.2 &    10.5 $\pm$  1.0 &    2.6 \\
 00 42 45.10 & +41 14 07.2 &   17.7 &    60.7 &     5.9 $\pm$  0.8 &    1.5 \\
 00 42 45.11 & +41 16 21.8 &   49.7 &   187.8 &    17.9 $\pm$  1.3 &    4.4 \\
 00 42 45.22 & +41 16 11.3 &   24.4 &    86.6 &     8.2 $\pm$  0.9 &    2.0 \\
 00 42 45.58 & +41 16 08.8 &   28.2 &   100.7 &     9.6 $\pm$  1.0 &    2.4 \\
 00 42 45.99 & +41 16 19.7 &   17.7 &    59.7 &     5.7 $\pm$  0.8 &    1.4 \\
 00 42 46.14 & +41 15 43.5 &   13.8 &    43.3 &     4.1 $\pm$  0.6 &    1.0 \\
 00 42 46.89 & +41 21 18.9 &    6.3 &    27.6 &     2.8 $\pm$  0.9 &    0.7 \\
 00 42 46.95 & +41 16 15.7 &  163.2 &   715.4 &    68.2 $\pm$  2.6 &   16.8 \\
 00 42 47.15 & +41 16 28.6 &  274.6 &  1290.6 &   122.9 $\pm$  3.4 &   30.4 \\
 00 42 47.17 & +41 14 13.2 &    4.7 &    13.3 &     1.3 $\pm$  0.4 &    0.3 \\
 00 42 47.28 & +41 11 58.4 &    8.0 &    37.1 &     3.8 $\pm$  0.7 &    0.9 \\
 00 42 47.44 & +41 15 07.8 &   22.8 &    70.9 &     6.9 $\pm$  0.8 &    1.7 \\
 00 42 47.87 & +41 15 50.2 &    7.3 &    23.2 &     2.2 $\pm$  0.5 &    0.5 \\
 00 42 47.87 & +41 15 33.1 &  207.6 &   838.5 &    80.8 $\pm$  2.8 &   19.9 \\
 00 42 48.48 & +41 25 21.9 &   49.2 &  1033.8 &   114.0 $\pm$  4.1 &   28.2 \\
 00 42 48.50 & +41 15 21.4 &  435.5 &  1901.4 &   183.7 $\pm$  4.2 &   45.4 \\
 00 42 48.69 & +41 16 24.7 &   12.3 &    49.1 &     4.7 $\pm$  0.7 &    1.2 \\
 00 42 49.13 & +41 17 42.3 &    5.1 &    15.6 &     1.5 $\pm$  0.4 &    0.4 \\
 00 42 49.20 & +41 18 16.0 &   35.4 &   129.4 &    12.4 $\pm$  1.1 &    3.1 \\
 00 42 51.28 & +41 16 39.7 &   21.7 &    74.2 &     7.1 $\pm$  0.8 &    1.8 \\
 00 42 51.67 & +41 13 03.3 &    9.0 &    35.8 &     3.6 $\pm$  0.7 &    0.9 \\
 00 42 52.01 & +41 31 07.6 &   57.7 &  4727.5 &   679.9 $\pm$ 10.6 &  167.9 \\
 00 42 52.30 & +41 17 34.7 &    7.1 &    22.9 &     2.2 $\pm$  0.5 &    0.5 \\
 00 42 52.50 & +41 15 40.1 &  403.3 &  2053.1 &   199.4 $\pm$  4.4 &   49.3 \\
 00 42 52.50 & +41 18 54.4 &  291.2 &  1914.5 &   187.3 $\pm$  4.3 &   46.3 \\
 00 42 52.58 & +41 13 28.1 &    4.2 &    13.7 &     1.4 $\pm$  0.5 &    0.3 \\
 00 42 53.49 & +41 25 49.9 &    8.8 &   183.4 &    20.6 $\pm$  2.3 &    5.1 \\
 00 42 54.91 & +41 16 03.3 &  341.8 &  1702.3 &   166.5 $\pm$  4.0 &   41.1 \\
 00 42 55.18 & +41 18 36.4 &   16.7 &    81.4 &     8.0 $\pm$  1.0 &    2.0 \\
 00 42 55.30 & +41 20 45.5 &    6.0 &    39.1 &     3.9 $\pm$  1.1 &    1.0 \\
 00 42 55.31 & +41 25 58.0 &   17.5 &   431.9 &    49.0 $\pm$  3.0 &   12.1 \\
 00 42 55.59 & +41 18 35.2 &   20.0 &    95.3 &     9.4 $\pm$  1.0 &    2.3 \\
 00 42 57.91 & +41 11 05.2 &   84.3 &   922.7 &    98.7 $\pm$  3.4 &   24.4 \\
 00 42 58.30 & +41 15 29.4 &   27.1 &   127.1 &    12.7 $\pm$  1.2 &    3.1 \\
 00 42 59.30 & +41 16 43.3 &   31.3 &   141.8 &    14.0 $\pm$  1.2 &    3.5 \\
 00 42 59.51 & +41 12 42.8 &    4.2 &    22.5 &     2.3 $\pm$  0.9 &    0.6 \\
 00 42 59.64 & +41 19 19.3 &  125.3 &  1046.3 &   104.9 $\pm$  3.3 &   25.9 \\
 00 42 59.86 & +41 16 06.0 &  180.8 &  1080.2 &   107.5 $\pm$  3.3 &   26.6 \\
 00 43 01.09 & +41 13 51.5 &    8.5 &    31.3 &     3.2 $\pm$  0.7 &    0.8 \\
 00 43 01.23 & +41 30 17.2 &    9.5 &   708.4 &    91.0 $\pm$  6.0 &   22.5 \\
 00 43 02.90 & +41 15 22.8 &  143.0 &   945.3 &    95.9 $\pm$  3.1 &   23.7 \\
 00 43 03.01 & +41 20 42.0 &    5.8 &    35.9 &     3.7 $\pm$  1.2 &    0.9 \\
 00 43 03.11 & +41 10 16.1 &   15.2 &   148.2 &    16.2 $\pm$  1.8 &    4.0 \\
 00 43 03.20 & +41 15 28.2 &   36.2 &   196.8 &    20.0 $\pm$  1.5 &    4.9 \\
 00 43 03.28 & +41 21 21.8 &   32.6 &   304.8 &    31.8 $\pm$  2.0 &    7.8 \\
 00 43 03.82 & +41 18 05.1 &   94.6 &   843.8 &    84.7 $\pm$  3.0 &   20.9 \\
 00 43 04.20 & +41 16 01.6 &   14.0 &    70.4 &     7.1 $\pm$  1.0 &    1.8 \\
 00 43 05.65 & +41 17 02.8 &  678.2 &  8796.6 &   891.7 $\pm$  9.5 &  220.3 \\
 00 43 06.98 & +41 18 10.5 &    8.9 &    49.3 &     5.0 $\pm$  1.2 &    1.2 \\
 00 43 07.47 & +41 20 21.0 &   17.5 &   182.8 &    19.1 $\pm$  1.7 &    4.7 \\
 00 43 08.45 & +41 12 47.4 &    8.9 &    95.8 &    10.2 $\pm$  1.3 &    2.5 \\
 00 43 09.82 & +41 19 00.8 &   78.5 &   831.2 &    85.9 $\pm$  3.1 &   21.2 \\
 00 43 10.59 & +41 14 51.4 &  180.5 &  1970.3 &   205.0 $\pm$  4.7 &   50.6 \\
 00 43 11.39 & +41 18 09.6 &    5.9 &    30.6 &     3.2 $\pm$  1.0 &    0.8 \\
 00 43 13.19 & +41 18 14.2 &   10.1 &    75.4 &     7.9 $\pm$  1.3 &    1.9 \\
 00 43 14.38 & +41 07 21.4 &   46.7 &  1233.7 &   148.0 $\pm$  5.1 &   36.6 \\
 00 43 16.06 & +41 18 41.5 &   11.5 &   124.8 &    13.2 $\pm$  1.8 &    3.3 \\
 00 43 18.54 & +41 09 50.0 &    5.9 &    60.3 &     7.0 $\pm$  2.6 &    1.7 \\
 00 43 18.92 & +41 20 16.7 &   16.3 &   258.4 &    28.2 $\pm$  2.8 &    7.0 \\
 00 43 20.99 & +41 17 48.4 &    7.9 &    91.8 &     9.9 $\pm$  1.9 &    2.4 \\
 00 43 24.79 & +41 17 29.0 &    4.5 &    69.3 &     7.6 $\pm$  1.8 &    1.9 \\
 00 43 27.85 & +41 18 30.5 &   22.5 &   370.8 &    41.2 $\pm$  3.0 &   10.2 \\
 00 43 29.07 & +41 07 48.4 &   21.7 &   597.9 &    74.4 $\pm$  4.8 &   18.4 \\
 00 43 32.51 & +41 10 40.9 &   16.1 &   347.8 &    41.3 $\pm$  4.8 &   10.2 \\
 00 43 34.33 & +41 13 24.6 &   32.7 &   806.1 &    93.5 $\pm$  4.5 &   23.1 \\
 00 43 37.11 & +41 14 43.6 &   42.5 &  1263.8 &   147.5 $\pm$  5.2 &   36.4 \\
 00 43 53.62 & +41 16 54.0 &   23.0 &   947.0 &   123.3 $\pm$  6.0 &   30.5 \\
\enddata

\tablecomments{Table~\ref{table:src} contains for each source detected
in our Chandra observation: RA and DEC -- the coordinates in J2000, S/N
- the source detection significance in $\sigma$, Counts -- the net
counts, Flux -- the photon flux, and Luminosity -- the estimated
luminosity based on the spectral model described in the text.  We note
that source names should be formed by appending the coordinates,
truncated to the first decimal of seconds in RA and whole seconds in
DEC, to the identifier CXOM31, e.g.\ the first source in the list is
CXOM31 J004150.4+411337.} \end{deluxetable}

\begin{deluxetable}{llclcc}
\tablecaption{Coincidences of Chandra sources with infrared
sources and globular clusters
\label{glob_ir}}
\tablewidth{0pt}
\tablehead{
  \colhead{Chandra source} & 
  \colhead{2Mass source} & \colhead{$\Delta({\rm X-IR})$} & 
  \colhead{GC} & \colhead{$\Delta({\rm X-GC})$} &
   \colhead{$\Delta({\rm GC-IR})$} }
\startdata
J004209.6+411745 & 0042094+411745 & 1.3 & MIT 140 & 0.8 & 0.5 \\
J004210.2+411510 & 0042102+411510 & 0.1 &         &     &     \\
J004212.1+411758 & 0042121+411758 & 0.4 & 078-140 & 0.6 & 0.5 \\
J004215.9+410115 & 0042158+410114 & 1.6 & 082-144 & 1.6 & 0.4 \\
J004218.6+411402 & 0042186+411402 & 0.0 & 086-148 & 0.2 & 0.4 \\
J004221.5+411419 & 0042215+411419 & 0.1 &         &     &     \\
J004226.0+411915 & 0042260+411914 & 0.2 & 096-158 & 0.5 & 0.3 \\
J004231.2+411938 & 0042312+411938 & 0.0 & MIT 192 & 0.6 & 0.3 \\
J004240.5+411033 & 0042406+411033 & 1.3 & 123-182 & 1.3 & 0.0 \\
J004241.4+411524 &                &     & MIT 213 & 1.5 &     \\
J004252.0+413107 & 0042519+413107 & 0.6 & 135-192 & 0.5 & 0.2 \\
J004255.5+411835 &                &     & 138-000 & 0.3 &     \\
J004259.6+411919 &                &     & 143-198 & 0.0 &     \\ 
J004259.8+411606 &                &     & 144-000 & 0.5 &     \\
J004301.2+413017 & 0043014+413017 & 2.4 & MIT 236 & 2.3 & 0.2 \\
J004302.9+411522 &                &     & 146-000 & 0.6 &     \\
J004303.2+412121 &                &     & 147-199 & 0.5 &     \\
J004303.8+411805 &                &     & 148-200 & 0.3 &     \\
J004307.4+412021 &                &     & 150-203 & 1.5 &     \\ 
J004310.5+411451 & 0043106+411451 & 0.5 & MIT 251 & 1.0 & 0.5 \\
J004314.3+410721 & 0043144+410721 & 0.4 & 158-213 & 0.6 & 0.1 \\
J004337.1+411443 & 0043372+411443 & 2.0 & MIT 299 & 2.3 & 0.3 \\
\enddata

\tablecomments{ Table~\ref{glob_ir} gives coincidences between Chandra
HRC sources and infrared sources and globular clusters. The table
contains: Chandra source -- the Chandra HRC source (CXOM31), 2Mass
source -- the 2mass infrared source, $\Delta({\rm X-IR})$ -- the
displacement between the Chandra source and the 2mass source in
arcseconds,  GC -- the name of the globular cluster, $\Delta({\rm
X-GC})$ -- the displacement between the Chandra source and the globular
cluster in arcseconds,  $\Delta({\rm X-IR})$ -- the displacement
between the globular cluster and the 2mass source in arcseconds. 
Globular clusters taken from \citet{barmby01} are listed as two sets of
three digits and those taken from \citet{magnier93} are listed as MIT
followed by three digits.} \end{deluxetable}


\end{document}